\documentclass[12pt]{article}
\usepackage{amssymb}
\usepackage[dvips]{epsfig}

\begin{document}

\author{A. E. R. Chumbes, A. E. O. Vasquez, M. B. Hott\thanks{%
marcelo.hott@pq.cnpq.br} \\
\\
UNESP Univ Estadual Paulista - Campus de Guaratinguet\'{a} - DFQ.\\
Av. Dr. Ariberto Pereira Cunha, 333\\
12516-410 Guaratinguet\'{a} SP Brasil}
\title{Fermion localization on a split brane}
\maketitle

\begin{abstract}
In this work we analyze the localization of fermions on a brane embedded in
five-dimensional, warped and non-warped, space-time. In both cases we use
the same nonlinear theoretical model with a non-polynomial potential
featuring a self-interacting scalar field whose minimum energy solution is a
soliton (a kink) which can be continuously deformed into a two-kink. Thus a
single brane splits into two branes. The behavior of spin 1/2 fermions
wavefunctions on the split brane depends on the coupling of fermions to the
scalar field and on the geometry of the space-time.
\end{abstract}

\section{Introduction}

The idea that our Universe can be realized inside a domain wall embedded in
a (4,1)-dimensional world \cite{rubakov1} has provided many creative ways to
solve the hierarchy of interactions problem \cite{hamed-dvali}-\cite%
{randall1} and the cosmological constant problem \cite{rubakov2} in worlds
with large extra dimension, without resorting to the compactification of the
extra dimension. It has also been shown \cite{randall2} that the effective
gravitational potential between two particles recovers the Newtonian
behavior, since one has localization of gravitons on a thin brane in
five-dimensional space-time with a warped geometry and the cosmological
constant is related to the brane tension. Localization of matter (spin-zero,
spin-1/2 and spin-3/2) in the Randall-Sundrun framework was shown to be
possible, under certain conditions over the brane tension \cite{gabadadze}.
The localization of spin 1/2 fermions on thin branes is due to a soliton,
\textit{via} the mechanism created by Jackiw and Rebbi \cite{jackiw-rebbi}
to demonstrate the phenomenon of fermion charge fractionization. By its
turn, the domain wall which we would live in, according to the scenario
proposed by Rubakov and Shaposhnikov \cite{rubakov1}, is the topological
defect provided by the very same soliton.

The Rubakov-Shaposhnikov scenario \cite{rubakov1} has been extended to
five-dimensional warped space-time by means of a self-interacting scalar
field, or a set of (self-)interacting scalar fields also coupled to gravity
\cite{varios1}. In those nonlinear models, also inspired on previous studies
on the stabilization of gravity fluctuations on domain walls in supergravity
theories \cite{cvetic1}, the thick branes solutions are minimum energy
configurations (in many cases they are Bogomol'nyi-Prasad-Sommerfield (BPS)
solutions) that separates the space in two patches with a peculiar
warp-factor whose asymptotic behavior is an anti-de Sitter (AdS$_{5}$) space.

In some models the minimum energy configurations are wide solitons \cite%
{melfo1}-\cite{dutra1}, also called double-wall. Double-wall configurations
also appear as solutions, degenerate in energy \cite{dutraamarohott}, of a
model with two interacting scalar fields \cite{bazeia2}-\cite{eto}, and they
have been called degenerate Bloch branes. The model used in \cite{bazeia2}
also comprises critical Bloch branes \cite{dutraamarohott} which are
analogues of the extreme domain walls described in \cite{cvetic1}. Such a
variety of solutions \cite{varios2} is due to an arbitrary constant of
integration of the equations of motion. The continuous deformation from a
single brane (domain wall) to a double-brane, up to an extreme one, by
varying the constant of integration, is similar to the phenomenon of brane
splitting discussed in \cite{campos1}, in analogy to first-order phase
transitions in condensed matter systems. Such a transition is usually
approached by using a sixth degree polynomial potential, as was done in \cite%
{campos1}, which is characterized by the increase of a disordered phase (a
wet phase in surface physics) or an increase of the thickness of the domain
wall (brane). As the temperature of the system approaches a certain limit,
one has the appearance of two interfaces between the disordered phase and
the ordered ones, that is, the formation of a double-wall. That phenomenon
has been called \textit{brane splitting }\cite{campos1}. As the temperature
of the system goes towards a critical one, a complete domination of the
disordered phase (complete wetting) might happen, which we interpret, in
brane worlds scenario, as the formation of an extreme brane. Such a phase
transition can also be described by effective non-polynomial potentials \cite%
{chumbeshott} constructed from the model with two scalar fields \cite%
{bazeia2}, \cite{varios2}, \cite{dutraamarohott} with the advantage of
having a model whose minimum energy solutions are the BPS ones.

The localization of fermions on a double-wall (double-brane) in warped
space-time has been studied in \cite{melfo2} and in \cite{gomes-cas}. It has
been shown that double-wall can localize massless fermions. Notwithstanding,
a close inspection on the behavior of the zero-mode eingenfunction reveals
that it is peaked just in the region between the branes and has tails
inside the branes, where the possibility of detecting massless fermions is
very small. It can be noticed that this behavior is due to the Yukawa
coupling, $\phi \bar{\Psi}\Psi $, of fermions to the soliton. We think that
the peak of the zero-mode eingenfunction should follows the trend of the
brane, that is, it should also split. In this work we show that this
behavior can be achieved by means of a convenient coupling of the fermion to
the scalar field which is reminiscent from $N=1$ supersymmetry (SUSY),
namely $W_{\phi \phi }\bar{\Psi}\Psi $, where $W_{\phi \phi }$ is the 2nd
derivative of the superpotential, of the field theory model, with respect to
the field taken at the BPS configuration.

We develop the calculations by using one of the models obtained in \cite%
{chumbeshott}, but they could also be done by using any other
nonlinear model which admits two-kink solutions for the scalar
field. We also analyze the localization of massless and massive
fermions in the Rubakov-Shaposhnikov framework \cite{rubakov1}, that
is, in a non-warped space-time. In this latter case one clearly sees
that the coupling $\phi \bar{\Psi}\Psi $ provides an effective
single well potential where the fermions would be trapped in, but
that well is just in the region between the branes, while the
coupling $W_{\phi \phi }\bar{\Psi}\Psi $ provides a double-well
potential with wells in the cores of the branes, as it should be. In
the case of flat space-time, the Numerov method is used to obtain
the massive localized modes, with emphasis on the SUSY inspired
coupling with which there is room to prepare a chiral mixed state of
quasi-degenerate massless and tiny mass fermion states. We also show
that such a mixed state can tunnel between the two branes with time
of tunneling inversely proportional to the tiny mass which, in its
turn, decreases as the distance between the walls increases. We
would like to warn the reader that the brane splitting is not
considered here as a dynamic process; in fact, the distance between
the walls is one of the parameters of the nonlinear potential that
could be seen as dependent on the temperature and the tunneling of
fermions is analyzed at a fixed distance between the walls, that is,
at a given temperature close to the critical temperature for the
formation of an extreme wall.

In the next section we show the main features of the nonlinear model coupled
to gravity in 5-dimensional space-time \cite{chumbeshott}. In the third
section we are concerned to the localization of massless fermions on a split
brane in warped space-time. In the fourth section we deal with the
localization and tunneling of massive fermions in flat 5-dimensional
space-time. \ A few remarks on fermion localization on double-walls are left
to the last section.

\section{The model, the double-brane and the warp factor}

The model we are going to deal with includes a self-interacting scalar field
coupled minimally to gravity with one extra space dimension, denoted by $r$.
The action that leads to Euler-Lagrange for the scalar field and Einstein
equations is given by%
\begin{equation}
S=\int d^{4}xdy\sqrt{|g|}\left( -\frac{1}{4}R+\frac{1}{2}g_{ab}\partial
^{a}\phi \partial ^{b}\phi -V(\phi )\right) ,  \label{eq1}
\end{equation}%
where $g\equiv \mathrm{Det}(g_{ab})$ and the metric is
\begin{equation}
ds^{2}=g_{ab}dx^{a}dx^{b}=e^{2A(r)}\eta _{\mu \nu }dx^{\mu }dx^{\nu }-dr^{2},%
\hspace{0.5in}a,b=0,...,4,  \label{eq2}
\end{equation}%
where $\eta _{\mu \nu }$ is the Minkowski metric and $e^{2A(r)}$ is the warp
factor, which is supposed to depend only on the extra dimension $r$. The
Greek indices run from $0$ to $3$.

We consider that the potential $V(\phi )$ can be written as
\begin{equation}
V(\phi )=\frac{1}{2}\left( \frac{dW(\phi )}{d\phi }\right) ^{2}-\frac{4}{3}%
\left( W(\phi )\right) ^{2},  \label{eq3}
\end{equation}%
\noindent In this case the BPS\ solution of the following first-order
differential equations
\begin{equation}
\frac{d\phi }{dr}=\pm \frac{dW(\phi )}{d\phi }\hspace{0.25in}\mathrm{and%
\hspace{0.25in}}\frac{dA}{dr}=\mp \frac{2}{3}W(\phi )  \label{eq4}
\end{equation}%
are also solutions of the second-order differential equations\ of motion in
the static limit.

By taking $W(\phi )$, which we call the superpotential, given by
\begin{equation}
W(\phi )=2\mu \left[ \phi \left( \frac{\phi ^{2}}{3}+f^{2}+\frac{b}{2}\sqrt{%
\phi ^{2}+f^{2}}\right) +\frac{bf^{2}}{2}\sinh ^{-1}\left( \frac{\phi }{f}%
\right) \right] ,  \label{eq5}
\end{equation}%
with $f=\sqrt{b^{2}-a^{2}}$ and $b<-a<0$, we have
\begin{equation}
\frac{dW(\phi )}{d\phi }={W}_{\phi }(\phi )=2\mu (\phi ^{2}+f^{2}+b\sqrt{%
\phi ^{2}+f^{2}}),  \label{eq6}
\end{equation}%
and the solutions of the first-order differential equations (\ref{eq4}) can
be found straightforwardly. The classical configuration for the scalar field
is given by
\begin{equation}
\phi =\pm a\frac{\sinh (2\mu ar)}{\cosh (2\mu ar)-b/f}~,  \label{eq7}
\end{equation}%
where the (lower)upper sign stands for (anti-)kink configuration. The
reference \cite{chumbeshott} has more details on the soliton profile, as
well as on the behavior of the warp factor in this model. As has been noted
in the second paper of reference \cite{varios2}, the solution (\ref{eq7})
can be conveniently written as%
\begin{equation}
\phi =\pm \frac{a}{2}[\tanh (\mu a(r+L))+\tanh (\mu a(r-L))],  \label{eq8}
\end{equation}%
where we have identified $b=-\cosh (2L)/\sqrt{\cosh (2L)^{2}-1}$. The above
expression can be seen as a merging of two solitons (two kinks), namely $%
\phi _{\pm }=\frac{a}{2}[\pm 1+\tanh (\mu a(r\mp L))]$ , which are localized
at $r=-L$ and $r=+L$.

In flat two-dimensional space-time one constructs such a scalar model as the
bosonic sector of a $N=1$ SUSY model with the potential given by only the
first contribution at the right hand side of the equation (\ref{eq3}). In
such a circumstance one can clearly see that the potential has two minima at
$\phi =\pm a$ and one additional minimum appears at $\phi =0$ as $b=-a$.
This transition in the potential is reflected in the behavior of the soliton
in (\ref{eq8}), which is continuously deformed from one kink into a two-kink
profile and finally in one of the solutions, $\phi _{\pm }$, for $b=-a$ (in
this critical case $L$ is replaced by an arbitrary constant of integration).
A first-order phase transition takes place. It is characterized by the
growing of the disordered phase with the formation of two interfaces
separating the disordered phase from the ordered ones, up to the complete
wetting in the critical limit. The phase transition can also be seen from
the behavior of the warp factor, discussed in \ \cite{chumbeshott}. One can
see that the warp factor separates the space along the extra dimension in
two similar regions, whose asymptotic behavior is an AdS$_{5}$ space and, as
$L$ increases, a double-brane structure is triggered at a specific value of $%
\ L$. Such a double-brane structure is evident from the peaks of the Ricci
scalar $R=-(8$ $d^{2}A/dr^{2}+20$ $(dA/dr)^{2})$ just at the points the core
of the branes are localized (see Figure 1 below). Moreover, in the limit $%
b=-a$ the branes are infinitely separated from each other, there being the
formation of an extreme brane. Such an extreme case is discussed in
references \cite{dutraamarohott} and \cite{dutracouceirohott} where the
brane is called a \textit{critical Bloch brane. }In the next sections we
focus only on the localization of massless fermions on single brane that
splits into two brane, that we also call double-brane or double-wall.

\section{Localization of fermions: warped space-time}

In this section we analyze the localization of massless fermions on the
brane described in the previous section by taking into account also the
curvature of the space. The problem is approached by searching normalized
solutions of the Dirac equation for fermions coupled with the scalar field
by a general Yukawa coupling $\bar{\Psi}F(\phi )\Psi $, where $F(\phi )$ is
a functional of the field $\phi (r)$ taken at the classical solution. The
functional form of $F(\phi )$, as well as the dependence of the metric on $r$%
, is crucial for the possible localization of massless fermions on the
brane, as we discuss below.

The fermion action defined as%
\begin{equation}
S_{fermion}=\int d^{5}x\sqrt{-g}~(\bar{\Psi}i\Gamma ^{a}D_{a}\Psi -\bar{\Psi}%
F(\phi )\Psi ),  \label{1}
\end{equation}%
leads to the Dirac equation

\begin{equation}
\lbrack i\Gamma ^{a}D_{a}-F(\phi )]\Psi =0.  \label{3}
\end{equation}%
The gamma matrices satisfy the algebra $\{\Gamma ^{a},\Gamma ^{b}\}=2g^{ab}$%
, and can be defined in the irreducible representation as $\Gamma ^{\mu
}=e^{-A(y)}\gamma ^{\mu }$, $\Gamma ^{5}=-i\gamma ^{5}$ in terms of the $%
4\times 4$ gamma matrices, $\gamma ^{a}$, in flat space-time. By means of
the definition of the covariant derivative $D_{a}=(\partial _{a}+\omega
_{a})\,\ $and the metric in (\ref{eq2}), one finds that \ the only
non-vanishing components of the connection $\omega _{a}$ are $\omega _{\mu }=%
\frac{1}{2}e^{A}(\partial _{r}A)\gamma _{\mu }\gamma _{5}$ and that the
Dirac equation turns out to be written as

\begin{equation}
\{i\gamma ^{\mu }\partial _{\mu }+e^{A(r)}\gamma ^{5}[\partial
_{r}+2\partial _{r}A(r)]-e^{A(r)}F(\phi )\}\Psi (x,r)=0.  \label{9}
\end{equation}%
To ease the separation of variables one can resort to the chiral
decomposition ($x$ stands for the four space-time coordinates)

\begin{equation}
\Psi (x,r)=\sum_{n}\psi _{Ln}(x)\alpha _{Ln}(r)+\sum_{n}\psi _{Rn}(x)\alpha
_{Rn}(r),  \label{10}
\end{equation}

\noindent where $\psi _{L(R)n}(x)$ are the chiral modes which satisfy $%
\gamma ^{5}\psi _{Ln}(x)=-\psi _{Ln}(x)$, $\gamma ^{5}\psi _{Rn}(x)=\psi
_{Rn}(x)$ and also the four-dimensional massive Dirac equations $i\gamma
^{\mu }\partial _{\mu }\psi _{L(R)n}=m_{n}\psi _{R(L)n}.$Then, we find the
following differential equations for the $r$-dependent scalar parts of the
spinor

\begin{equation}
\alpha _{Rn}^{\prime }+[2A^{\prime }-F(\phi )]\alpha
_{Rn}=-m_{n}e^{-A}\alpha _{Ln},  \label{13a}
\end{equation}%
and%
\begin{equation}
\alpha _{Ln}^{\prime }+[2A^{\prime }+F(\phi )]\alpha _{Ln}=m_{n}e^{-A}\alpha
_{Rn},  \label{13b}
\end{equation}%
where the prime stands for the first-derivative with respect to $r$. The
equations above can be put in a more familiar form, namely

\begin{equation}
R_{n}^{\prime }-F(\phi )R_{n}=-m_{n}e^{-A}L_{n}~\mathrm{and~\ }L_{n}^{\prime
}+F(\phi )L_{n}=m_{n}e^{-A}R_{n},  \label{14}
\end{equation}%
by using the redefinitions $\alpha _{Rn}(r)=e^{-2A(r)}R_{n}(r)$ and $\alpha
_{Ln}(r)=e^{-2A(r)}L_{n}(r)$.

Equations (\ref{14}) are equivalent to the equations for the components of a
spinor describing a massless fermion in 1+1 dimensions subject to a mixing
of scalar and vector potentials. The time-independent equation for a fermion
under such potentials can be written as $H\psi (r)=E\psi (r)$, with the
Dirac Hamiltonian given (in natural units) by $H=\sigma _{2}p+V_{s}(r)\sigma
_{1}+V_{v}(r)$, where $p=-id/dr$ is the momentum operator, $\sigma _{1}$ and
$\sigma _{2}$ are the two off-diagonal Pauli matrices, $V_{s}(r)=$ $-F(\phi )
$ (note that $F(\phi )$ is a function of $r$) is the scalar potential and $%
V_{v}(r)=m_{n}e^{-A(r)}$ is the vector potential. Analogously, $L_{n}(r)$
and $R_{n}(r)$ play the role of the upper and lower components, for the
fermion zero-mode in 1+1 dimensions. As a matter of fact, the scalar
potential can be seen as a position-dependent fermion mass. As one knows,
many examples of such systems were already solved in the literature \cite%
{castro}, but only a few potentials may support bound states.

In the brane world scenario, by its turn, massive ($m_{n}\neq 0$) as well as
massless ($m_{n}=0$) fermions might be localized inside the brane, depending
on the functional coupling $F(\phi )$. The issue of localization of massive
modes on the brane is not going to be discussed in this section, even
though, we would like to remark that the analysis of the possible massive
spectrum of bound states is not as simple as the search for localized
massless modes \cite{dutracouceirohott}, \cite{koley}. Moreover,
normalizable $\alpha _{R(L)n}(r)-$functions is guaranteed if they satisfy
the ortonormalization relations

\begin{eqnarray}
\int_{-\infty }^{\infty }e^{3A(r)}\alpha _{Ln}(r)\alpha _{Lm}(r)dr
&=&\int_{-\infty }^{\infty }e^{3A(r)}\alpha _{Rn}(r)\alpha _{Rm}(r)dr=\delta
_{nm},  \nonumber \\
&&  \nonumber \\
\int_{-\infty }^{\infty }e^{3A(r)}\alpha _{Ln}(r)\alpha _{Rm}(r)dr &=&0.
\label{15}
\end{eqnarray}%
From these relations one can note that the warp factor is crucial to
determine the localization of whatever mode.

Particularly, for the massless mode one finds that

\begin{eqnarray}
\alpha _{R_{0}}(r) &=&N_{R_{0}}\exp [-2A(r)+\int^{r}F(r^{\prime })dr^{\prime
}]\;  \nonumber \\
\alpha _{L_{0}}(r) &=&N_{L_{0}}\exp [-2A(r)-\int^{r}F(r^{\prime })dr^{\prime
}]~,  \label{16}
\end{eqnarray}%
where we have used $F(r)=F(\phi (r))\ $and $N_{R(L)~_{0}}$ are constants of
normalization that are found according to the normalization relations%
\begin{equation}
\left\vert N_{R_{0}}\right\vert ^{2}\int_{-\infty }^{\infty
}e^{-A(r)+2\int^{r}F(r^{\prime })dr^{\prime }}dr=\left\vert
N_{L_{0}}\right\vert ^{2}\int_{-\infty }^{\infty
}e^{-A(r)-2\int^{r}F(r^{\prime })dr^{\prime }}dr=1.  \label{17}
\end{equation}%
\qquad \qquad \qquad\ \qquad \qquad \qquad \qquad

One can note that the normalization conditions are determined by the
asymptotic behavior of the integrands in the expressions above. In most of
the examples dealing with thick branes the factor $\exp [-A(r)]$ goes
asymptotically to infinity given the asymptotic behavior of the warp factor,
that is $\exp [2A(r\rightarrow \pm \infty )]\rightarrow 0$, hence one has to
choose $F(\phi (r))$ in such a way that the decreasing of $\exp [\pm
\int^{r}F(r^{\prime })dr^{\prime }]$ is faster than\ the increasing of $\exp
[-A(r)]$, but this choice does not guarantee the normalization of both
chiralities, because of the different signs, $\pm $, in the exponent, that
is, if $\alpha _{R_{0}}(r)$ is normalizable, $\alpha _{L_{0}}(r)$ is not and
\textit{vice-versa}.\textit{\ }Moreover, the factor $\exp [-2A(r)]$ is
symmetric around the core of the brane whereas the kink solution is odd
under the reversion of the $r$ coordinate around the core of the brane,
which is usually chosen at $r=0$, hence one has to choose $F(\phi )$ as an
odd function on $r$, in order to have the normalizable chiral mode even in $%
r $ and with a peak on the brane.\qquad

We have analyzed the behavior of the massless chiral modes for two cases,
namely: $F(\phi )=\eta \phi (r)$ and $F(\phi )=-\eta W_{\phi \phi }$, where $%
\eta >0~$is a coupling constant and $W_{\phi \phi }$ is the
second-derivative of the superpotential (\ref{eq5}) with respect to $\phi $
taken at the two-kink configuration in (\ref{eq7}). The first case is the
simplest Yukawa coupling of fermions to a scalar real field, while the
second one is inspired on the coupling of fermions and bosons within a $N=1$
SUSY model, which is also considered in the next section. This last
functional coupling is the one which provides the correct localization on
the brane. In fact, the simplest Yukawa coupling also provides a localized
massless left-handed mode, $\alpha _{L_{0}}(r)$, but it does not follows the
brane splitting, that is, while the double-wall is formed, the peak of the
wavefunction is midway between the two walls, there being a small
probability density to find such a mode on the core of the walls themselves.
On the other hand, in the case $F(\phi )=-\eta W_{\phi \phi }$, one has $%
\alpha _{L_{0}}(r)~$with peaks on the branes, signalizing a great
probability for the massless left-handed mode to be found just on the branes
and a small probability to be found on the bulk and between the walls. Those
behaviors can be seen from Figure 2, where the profiles of $\alpha
_{L_{0}}(r)$ are shown for different values of $L$. Figure 2 should be
confronted with Figure 1.

\section{Localization of fermions: flat space-time}

In this section we adopt the usual analysis to find fermion bound states
under the action of the scalar field whose classical configuration is given
in (\ref{eq7}). Particularly, we focus on the scenario proposed in \cite%
{rubakov1}, that is, a brane (or domain wall) immersed in a five-dimensional
flat space-time. The action for the fermion field is given in (\ref{1}) with
$g_{ab}=\eta _{ab}$ the Mikowskian metric, $D_{a}\equiv \partial _{a}$ and
the irreducible form of the gamma matrices, $\Gamma ^{\mu }=\gamma ^{\mu }$,
is going to be used. The chiral decomposition (\ref{10}) can also be used to
separate the four space-time variables, $x^{\mu }$, from the variable $r$.
Now, the $r$-dependent functions appearing in the chiral decomposition obey
the following equations

\begin{equation}
\alpha _{Rn}^{\prime }-F(\phi )\alpha _{Rn}=-m_{n}\alpha _{Ln},  \label{18}
\end{equation}%
and%
\begin{equation}
\alpha _{Ln}^{\prime }+F(\phi )\alpha _{Ln}=m_{n}\alpha _{Rn}.  \label{19}
\end{equation}

Particularly, for the massless mode one finds

\begin{eqnarray}
\alpha _{R_{0}}(r) &=&N_{R_{0}}\exp [+\int^{r}F(r^{\prime })dr^{\prime }]\;
\nonumber \\
\alpha _{L_{0}}(r) &=&N_{L_{0}}\exp [-\int^{r}F(r^{\prime })dr^{\prime }]~.
\label{20}
\end{eqnarray}%
Now, the normalization of the massless modes depends on the asymptotic
behavior of $\ \int^{r}F(r^{\prime })dr^{\prime }$ only, hence one usually
has a unpaired (isolated) chiral (left-handed or right-handed) zero-mode,
which is the main feature for having fermion number fractionization, as
shown in \cite{jackiw-rebbi}.

We again have analyzed the behavior of the massless modes by setting $F(\phi
)=\eta \phi (r)$ and $F(\phi )=-\eta W_{\phi \phi }$, with $\eta >0$. In the
first case we have%
\begin{equation}
\alpha _{R_{0}}(r)=0~\mathrm{and~}\alpha _{L_{0}}(r)=N_{L_{0}}(\mathrm{Cosh}%
2L+\mathrm{Cosh}2r)^{-\eta /2}.  \label{21}
\end{equation}%
As in the previous section, the function $\alpha _{L_{0}}(r)~$is symmetric
in $r$ and has a peak at $r=0$, hence the massless left-handed mode is
localized in the region between the branes with a very small probability
density at the core of the branes.

For $F(\phi )=-\eta W_{\phi \phi }$ one finds%
\begin{equation}
\alpha _{R_{0}}(r)=0~\mathrm{and~}\alpha _{L_{0}}(r)=N_{L_{0}~}\left(
\mathrm{sech}^{2}(r+L)+\mathrm{sech}^{2}(r-L)\right) ^{\eta }.  \label{22}
\end{equation}%
From Figure 3 we can note that $\alpha _{L_{0}}(r)$ is symmetric in $r$, has
no nodes and exhibits peaks at the cores of the branes. One can also note
that the probability density to find the left-handed massless mode in the
midway between the branes decreases as $L$ increases.

The adequate behavior of the massless mode is sufficient enough to consider
the functional coupling $F(\phi )=-\eta W_{\phi \phi }$ as very convenient
and has motivated us to analyze the consequences of such a coupling on the
localization of massive modes on the split brane in flat space-time. As one
knows, equations (\ref{18}) and (\ref{19}) can be decoupled to two
second-order differential equations, namely

\begin{eqnarray}
-\alpha _{Rn}^{\prime \prime }+U_{R}(r)\alpha _{Rn} &=&m_{n}^{2}\alpha _{Rn},
\nonumber \\
-\alpha _{Ln}^{\prime \prime }+U_{L}(r)\alpha _{Ln} &=&m_{n}^{2}\alpha
_{Ln}~,  \label{23}
\end{eqnarray}%
where $U_{R}(r)=(\eta W_{\phi \phi })^{2}-\eta W_{\phi \phi
}^{\prime }$ and $U_{L}(r)=(\eta W_{\phi \phi })^{2}+\eta W_{\phi
\phi }^{~\prime }$ in the case $F(\phi )=-\eta W_{\phi \phi }$. It
is also known that the equations above are time-independent
Schr\"{o}dinger equations, whose corresponding Hamiltonians are
superpartners of each other, that is, one has a quantum mechanics
supersymmetry. This is formally true whatever is the functional
coupling $F(\phi )$, but in the case considered here such
supersymmetry seems to be a reflection of a supersymmetry at the
fundamental level. In other words, one can note that $r$-dependent
part of the excitations of the scalar field (\textit{branons}),
picked up to quadratic terms on the fundamental Lagrangian density
in flat space-time (now with $V(\phi )=W_{\phi }^{2}/2$), obeys a
time-dependent Schr\"{o}dinger equation similar to the one obeyed by
$\alpha _{Ln}(r)$ with $\eta =1$ , that is, with an effective
potential given by $U_{eff}(r)=(W_{\phi \phi })^{2}+W_{\phi \phi }^{\prime }$%
, resulting identical mass spectrum for \textit{branons }(bosonic
excitations)\textit{\ }and fermions .

In Figure 4 it is shown the form of the potentials $U_{R}(r)$ and $U_{L}(r)$
for a specific value of $L$ and $\eta =1$. For values of $L$ close to zero, $%
U_{L}(r)$ is a single well potential, which starts to be deformed into a
double-well potential as $L$ approaches to a critical value $L_{c}$, that is
determined by the condition $U_{L}^{\prime \prime }(r=0)=0$; for $%
L\gtrapprox L_{c}$, dimples are observed around $r=\pm L$, and a remarkable
double-well is observed for the value of $L=l$ corresponding to $U_{L}(r=0)=0
$. It worth mentioning that, although the bottom of the double-well is
slightly raised as $L$ increases, the width of the double-well potential
also increases, signalizing the possible entrapment of a massive state,
besides the, always present, massless one. This possible appearance of a
massive bound state can also be seen from the behavior of $U_{R}(r)$, which
is a single well potential whose bottom is above zero for  $L<l$ and equals
to zero for $L=l$, that is, the deepness and width of $U_{R}(r)$ increase as
$L$ increases.

One can also observe that $U_{L}(r)=2(2-3\mathrm{sech}^{2}r)$ and $%
U_{R}(r)=2(2-\mathrm{sech}^{2}r)$ for $L=0$ and $\eta =1$; such that the
first potential admits two bound states and the later admits only one bound
state. The fundamental state of \ $U_{L}(r)$ for $L=0$ and $\eta =1$ is $%
\alpha _{L_{0}}(r)\simeq \mathrm{sech}^{2}r$, whereas the first excited
state is $\alpha _{L_{1}}(r)$ $\simeq \mathrm{sech}r~\tanh r$ and the
fundamental state of $U_{R}(r)$\ for $L=0$ and $\eta =1$ is $\alpha
_{R_{1}}(r)\simeq \mathrm{sech}r$. \ Moreover, from the expression (\ref{22}%
) with $\eta =1$, one can construct \ an antisymmetric function as $\alpha
_{L_{1}}(r)$ $\sim \mathrm{sech}^{2}(r+L)-\mathrm{sech}^{2}(r-L)$ as an
approximate expression for the first excited state of $U_{L}(r)$ when $L>>l$%
. This approximation for the first excited state is commonly used to
approach the discrete spectrum of double well potentials \cite{merz}, \cite%
{sukhatme}.

We have used the results described above, together with the Numerov method
to analyze the behavior of $\alpha _{L_{1}}(r)$, $\alpha _{R_{1}}(r)$ and
the eigenvalue $m_{1}^{2}~$at intermediary values of $L$. Those behaviors
are shown in Figures 5 and 6. From them one can see that $\alpha _{R_{1}}(r)$
is mainly distributed on the region between the branes, hence there is a
small probability for the massive right-handed mode being observed inside
the wells where the Universe(s) would be realized, whereas the probability
density associated with $\alpha _{L_{1}}(r)$ is pronounced just on the cores
of the branes. In summary, at least one massive left-handed mode is
localized on the branes. The eigenvalue $m_{1}^{2}$ decreases smoothly as $L$
increases, that is an expected result when one is dealing with double-well
potentials in non-relativistic quantum mechanics. That way, one can
construct a mixed left-handed massive state with both, the fundamental and
the first excited state, which are quasi-degenerate for very large values of
$L$, namely

\begin{equation}
\Psi _{L,mix}(t,r)=N\left( \alpha _{L_{0}}(r)+\alpha
_{L_{1}}(r)e^{-im_{1}t}\right) \chi _{L},  \label{24}
\end{equation}%
where $\chi _{L}$ is a constant spinor which satisfies $\gamma ^{5}\chi
_{L}=-\chi _{L}$ . We have tried to be cautious when proposing this mixed
state, since we are assuming that there is a rest reference frame for the
particle in such a mixed state. With this reasoning, the Dirac equation $%
i\gamma ^{\mu }\partial _{\mu }\psi _{L,mix}=m_{n}\psi _{R}$ is satisfied,
since $m_{1}\gamma ^{0}\chi _{L}e^{-im_{1}t}=m_{1}\chi _{R}e^{-im_{1}t}$ ($%
\gamma ^{5}\chi _{R}=\chi _{R}$) and there is no right-handed massless
state, neither inside nor outside the branes. The probability density
associated with this mixed state is given by%
\begin{equation}
\rho (r,t)=|N|^{2}\left[ \alpha _{L_{0}}(r)^{2}+\alpha
_{L_{1}}(r)^{2}+2\alpha _{L_{0}}(r)\alpha _{L_{\ 1}}(r)\cos (m_{1}t)\right] ,
\label{25}
\end{equation}%
which is an oscillating probability density with period of oscillation $%
T=h/m_{1}c^{2}$, such that a tiny mass implies a long tunneling time. In
this scenario the fermion tunnels from one brane to the other, being likely
found on both branes, but not simultaneously. As $L$ increases, other
massive localized states can be realized inside the branes. In fact, we have
found numerically that there is room to one more massive state in the
present case. The appearance of a tower of localized massive states is very
dependent on the deepness and width of the double-well effective potential $%
U_{L}(r)$, that is ultimately dependent on the superpotential $W(\phi )$
whose classical solution is a kink that can be continuously deformed into
two-kink solution . In the next section we comment on the construction of
such models.

\section{Conclusions}

We have studied the mechanism that leads to the localization of \ massless
fermions on a split brane in the cases of warped and flat geometries. The
brane is immersed in a five-dimensional space-time and is defined by the
behavior of a scalar field coupled with gravity in the case of warped
space-time. The nonpolynomial potential of the self-interacting scalar field
which generates the split brane was introduced before in reference \cite%
{chumbeshott}, but any other model which has deformable solitons as minimal
energy configurations could be used as well, for example, a convenient $\phi
^{6}$ polynomial potential \cite{christlee}

The case of flat geometry is more manageable than the case of warped
geometry, not only because we obtain the localized modes in a simple way, \
but also, and mainly, because it allows us to understand why the coupling of
fermions with the scalar field should be chosen in such a way that a
supersymmetry in a fundamental level is realized. The most convenient
coupling, which preserves the supersymmetry, also leads to the expected
behavior of the fermion wavefunctions, that is, the ground-state
wavefunction follows the brane splitting. As the brane splits into two
branes the wavefunction is also split, with peaks at the cores of the two
branes. Although the simplest Yukawa coupling of fermions to scalar fields
also provides localization of massless fermions, the corresponding massless
fermion wavefunction present a peak just between the two branes, such that
the observation of massless fermions is suppressed inside the branes. Hence,
one concludes the supersymmetry inspired coupling as the most adequate one
to study\ the entrapment of fermions by a split brane.

We also note that eventual massive localized states are more
difficult to be found in the case of warped geometry than in the
case of flat geometry. In the later case we have also discussed the
tunneling of \ a massive fermion between the branes, which lasts
until the branes are infinitely separated from each other. That
critical limit is described by another model, obtained from the
starting one by taking the limit $b\rightarrow -a$ in (\ref{eq6}),
that is $V(\phi )=2\mu ^{2}(\phi ^{2}-a|\phi |)^{2}$, whose solution
is one of the single kinks $\phi =\frac{a}{2}[\pm 1+\tanh (\mu
a(r\mp r_{0}))]$, with $r_{0}$ a reference point  where the core of
the defect is localized. In this case we found that the energy gap
between the first localized states is large enough such that they
can not be considered as quasi-degenerate. It can also be observed
that the number of massive localized states depends on the deepness
and width of the quantum mechanics effective potentials, which are
defined by the field theory model one chooses to deal with. We have
mentioned above that others models whose classical solution exhibit
a two-kink profile (split brane) can afford a tower of localized
massive fermions states. A class of such models, called deformed
models, have been proposed \cite{bazeia}, as deformation of others
known models. We notice that those models can also be constructed,
together with new ones, from the deformation of zero modes
excitations of well know models whose classical solutions are single
kinks. This proposal is being analyzed in more detail and will be
reported later elsewhere.

\bigskip

\textbf{Acknowledgements: } The authors are grateful to CNPq and CAPES for
the financial support. MBH thanks to A. de Souza Dutra and J.M. Hoff da
Silva for many discussions on questions concerning the brane worlds
scenarios.

\newpage
\begin{figure}[tbp]
\begin{center}
\begin{minipage}{20\linewidth}
\epsfig{file=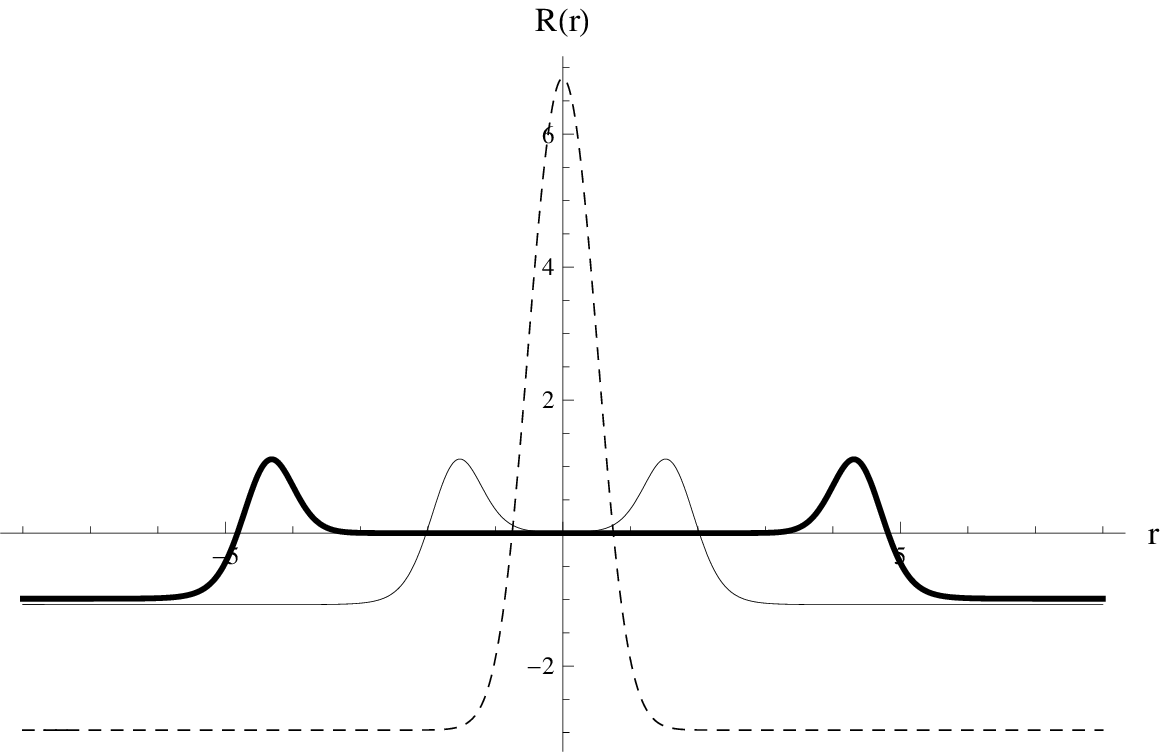}
\end{minipage}
\end{center}
\caption{Ricci scalar for $L=0.01$ (dashed line), $L=1.6$ (thin solid line)
and $L=4.5$ (thick solid line) evidences the formation of a double-wall
structure as $L$ increases.}
\label{fig:Fig.1}
\end{figure}

\begin{figure}[tbp]
\begin{center}
\begin{minipage}{20\linewidth}
\epsfig{file=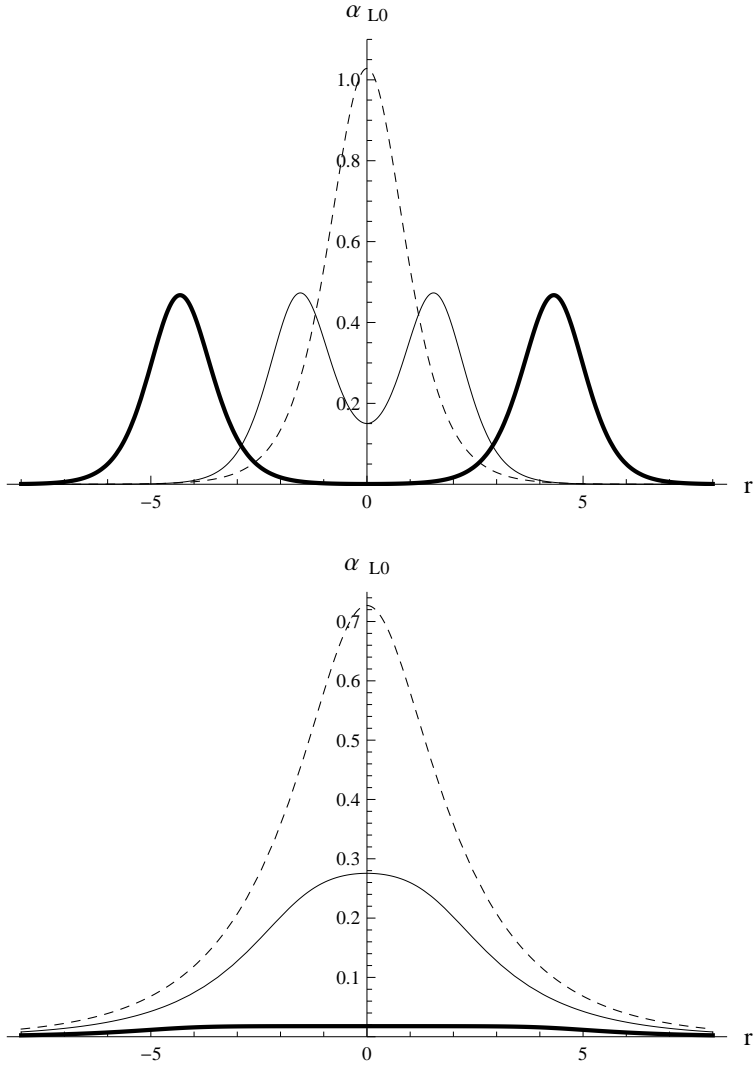}
\end{minipage}
\end{center}
\caption{$\protect\alpha _{L_{\ 0}}(r)$ (warped geometry) in the cases $F(%
\protect\phi )=\protect\phi (r)$ (upper) and $F(\protect\phi )=-W_{\protect%
\phi \protect\phi }$ (lower), for $L=0.01$ (dashed line), $L=1.6$ (thin
solid line) and $L=4.5$ (thick solid line).}
\label{fig:Fig.2}
\end{figure}

\begin{figure}[th]
\begin{center}
\includegraphics[width=15cm]{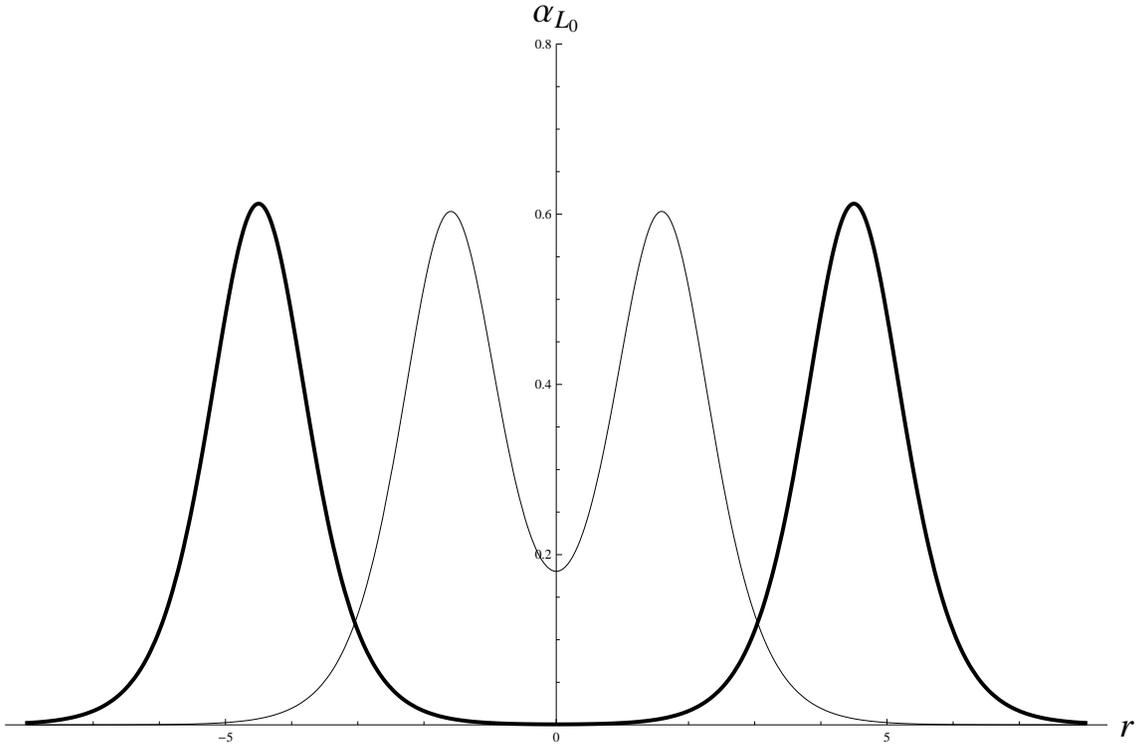} 
\end{center}
\caption{$\protect\alpha _{L_{\ 0}}(r)$ (flat space-time) in the case $F(%
\protect\phi )=-W_{\protect\phi \protect\phi }$, for $L=1.6$ (thin solid
line) and $L=4.5$ (thick solid line).}
\label{fig:Fig.3}
\end{figure}

\begin{figure}[tbp]
\begin{center}
\epsfig{file=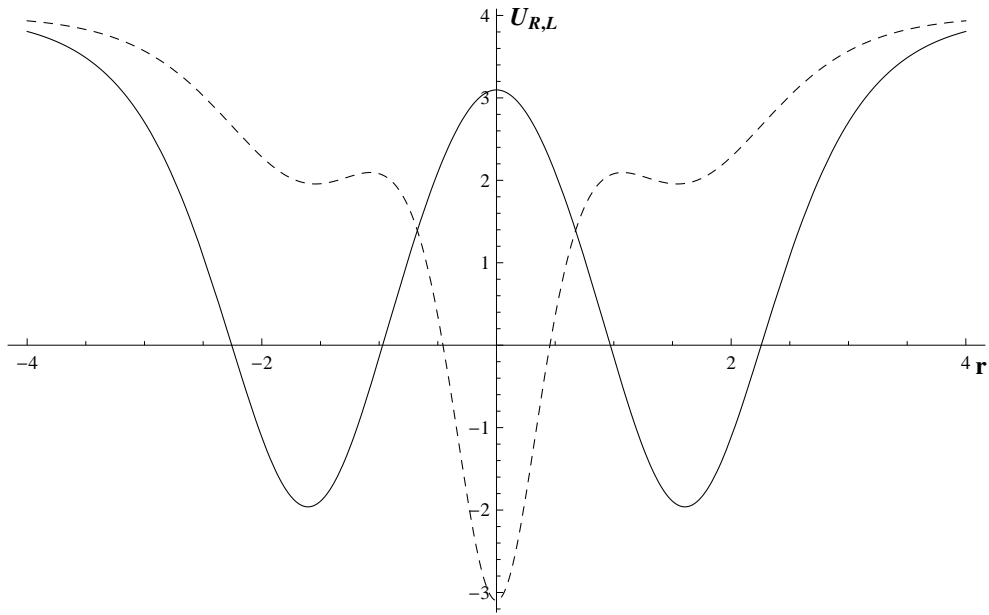} 
\end{center}
\caption{Effective potentials of equations (\protect\ref{22}) with $L=1.5$. $%
U_{L}(r)$ (solide line), $U_{R}(r)$ (dashed line).}
\label{fig:Fig.4}
\end{figure}

\begin{figure}[th]
\begin{center}
\includegraphics[width=15cm]{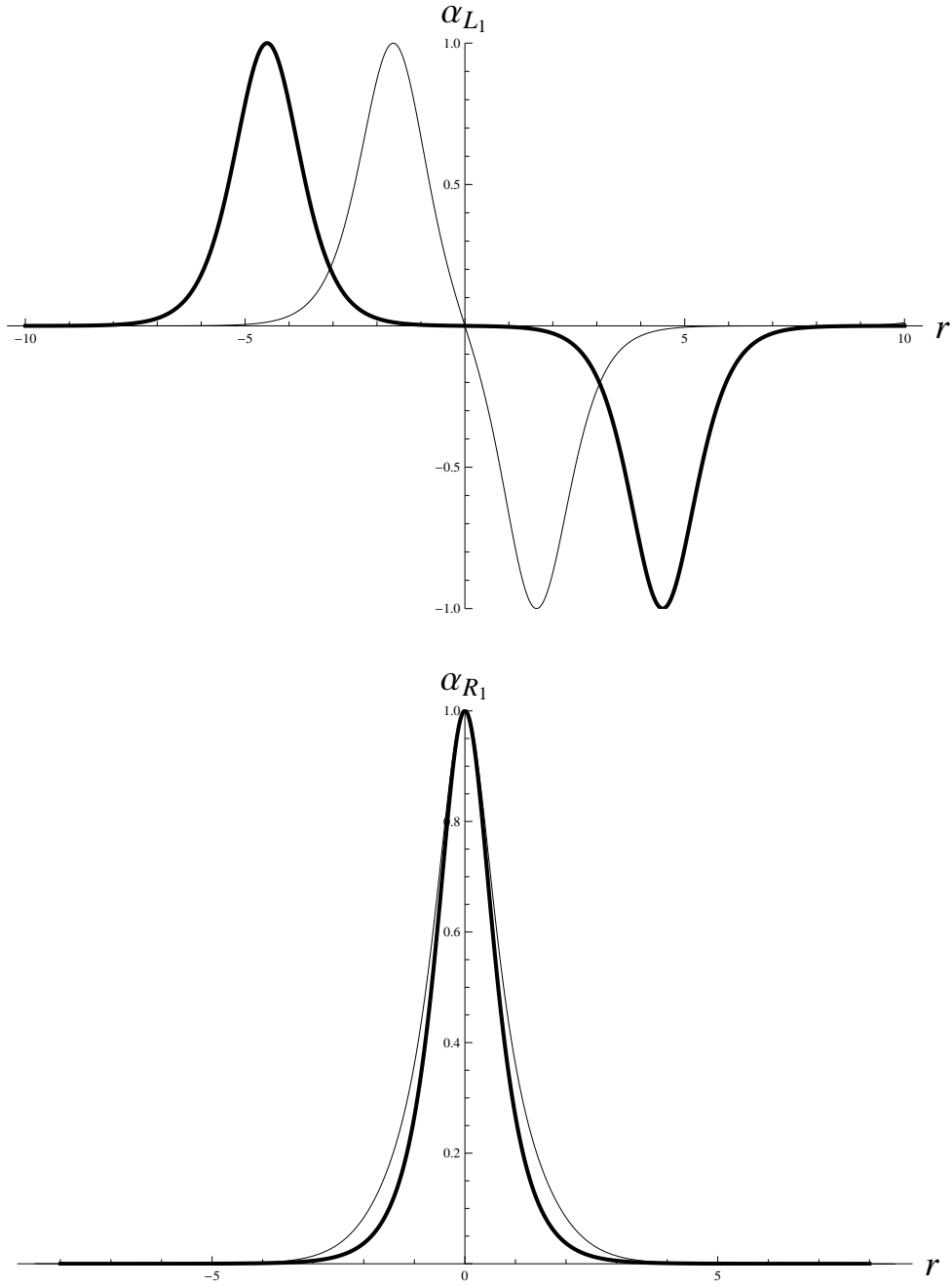} 
\end{center}
\par
\vspace{-0.2cm}
\caption{$\protect\alpha_{L1}(r)$ (upper) and $\protect\alpha_{R1}(r)$
(lower) in the case $F(\protect\phi )=-W_{\protect\phi \protect\phi }$, for $%
L=1.6$ (thin solid line) and $L=4.5$ (thick solid line), in flat space-time}
\label{fig:Fig.5}
\end{figure}

\begin{figure}[tbp]
\begin{center}
\epsfig{file=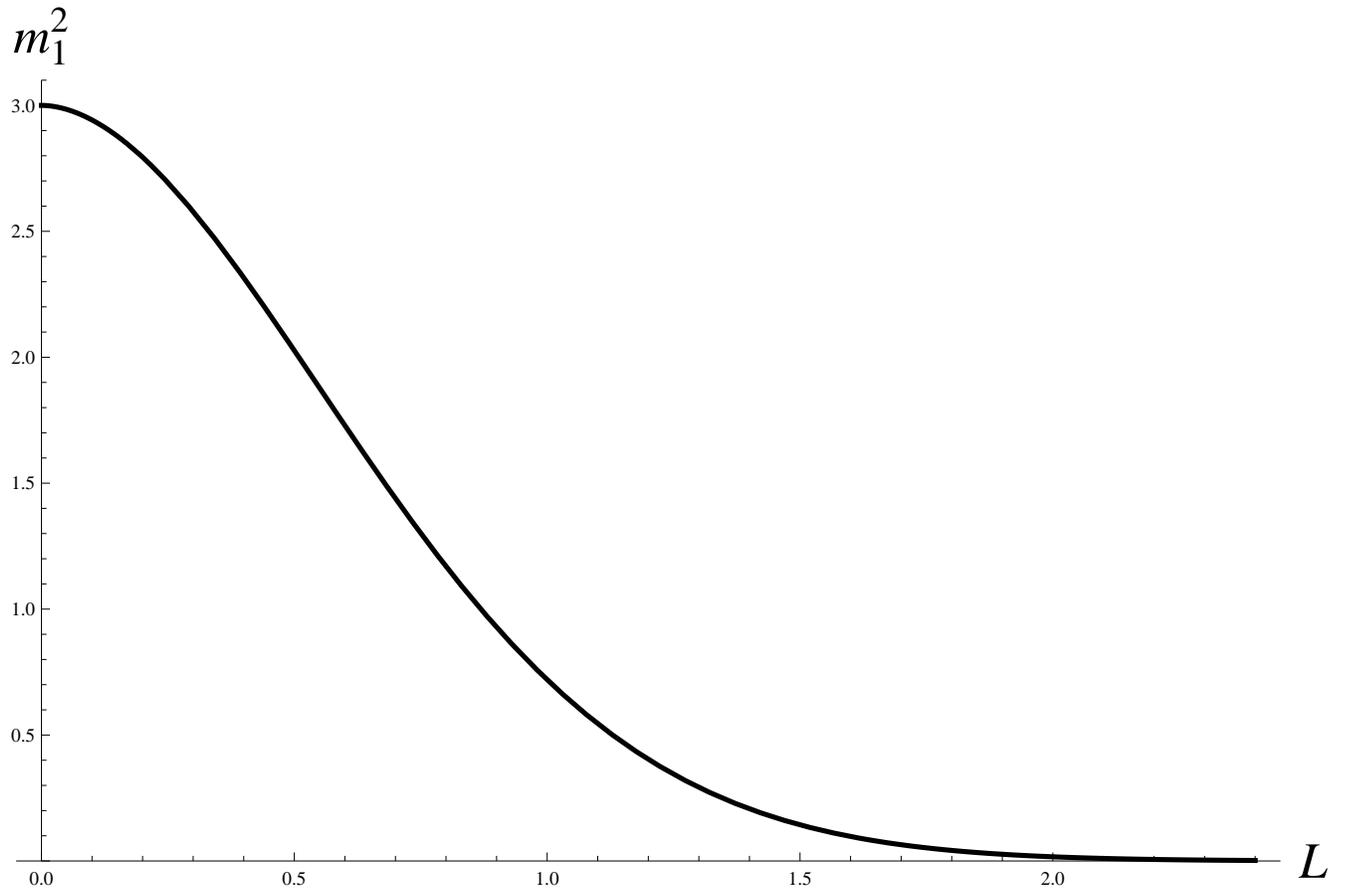} 
\end{center}
\caption{The eigenvalue of the first excited state in (\protect\ref{23})
against $L$.}
\label{fig:Fig.6}
\end{figure}

\end{document}